\documentclass[10pt]{article}
\usepackage{graphicx}

\def\Title#1{\begin{center} {\Large #1 } \end{center}}
\def\Author#1{\begin{center}{ \sc #1} \end{center}}
\def\Address#1{\begin{center}{ \it #1} \end{center}}

\newcommand\pubblock{\rightline{\begin{tabular}{l} Proceedings of the Fifth Annual LHCP\\ \pubnumber\\
         \pubdate  \end{tabular}}}

\newenvironment{Abstract}{\begin{quotation} \begin{center} 
             \large ABSTRACT \end{center}\bigskip 
      \begin{center}\begin{large}}{\end{large}\end{center} \end{quotation}}

\newenvironment{Presented}{\begin{quotation} \begin{center} 
             PRESENTED AT\end{center}\bigskip 
      \begin{center}\begin{large}}{\end{large}\end{center} \end{quotation}}

\def\beq{\begin{equation}}
\def\eeq#1{\label{#1}\end{equation}}
\def\eeqn{\end{equation}}

\def\beqa{\begin{eqnarray}}
\def\eeqa#1{\label{#1}\end{eqnarray}}
\def\eeqan{\end{eqnarray}}

\let\bar=\overbar

\def\Dslash{\not{\hbox{\kern-4pt $D$}}}
\def\dslash{\not{\hbox{\kern-2pt $\del$}}}

\def\msb{{\bar{\ssstyle M \kern -1pt S}}}

\usepackage{color}

 \newcommand\pubnumber{ }

\newcommand\pubdate{\today}

\def\affiliation{
On behalf of the CMS Collaboration, \\
CERN}

\begin{document}

\large
\begin{titlepage}
\pubblock

\vfill
\Title{ THE CMS HGCAL DETECTOR FOR HL-LHC UPGRADE }
\vfill

\Author{ ARABELLA MARTELLI }
\Address{\affiliation}
\vfill
\begin{Abstract}

The High Luminosity LHC (HL-LHC) will integrate 10 times more luminosity than the LHC, 
posing significant challenges for radiation tolerance and event pileup on detectors, 
especially for forward calorimetry, and hallmarks the issue for future colliders. 
As part of its HL-LHC upgrade program, the CMS collaboration is designing a High Granularity Calorimeter 
to replace the existing endcap calorimeters. 
It features unprecedented transverse and longitudinal segmentation for both electromagnetic (ECAL) and hadronic (HCAL) compartments. 
This will facilitate particle-flow calorimetry, where the fine structure of showers can be measured 
and used to enhance pileup rejection and particle identification, whilst still achieving good energy resolution. 
The ECAL and a large fraction of HCAL will be based on hexagonal silicon sensors of 0.5~-~1~cm$^{2}$ cell size, 
with the remainder of the HCAL based on highly-segmented scintillators with SiPM readout. 
The intrinsic high-precision timing capabilities of the silicon sensors will add an extra dimension to event reconstruction, 
especially in terms of pileup rejection. An overview of the HGCAL project is presented, covering motivation, 
engineering design, readout and trigger concepts, and performance (simulated and from beam tests).

\end{Abstract}
\vfill

\begin{Presented}
The Fifth Annual Conference\\
 on Large Hadron Collider Physics \\
Shanghai Jiao Tong University, Shanghai, China\\ 
May 15-20, 2017
\end{Presented}
\vfill
\end{titlepage}
\def\thefootnote{\fnsymbol{footnote}}
\setcounter{footnote}{0}
%

\normalsize 


\section{Introduction}
The high luminosity phase of the LHC (HL-LHC), expected to start its operation in about ten years,
will integrate 10 times more luminosity than the LHC, with the aim of pushing forward the demanding 
physics program of Phase II~\cite{EUstrategy2013}.
The high radiation and high pileup expected are major challenges for
the current detectors, which will be upgraded to maintain excellent performance even in the harsh HL-LHC environment.
As part of the HL-LHC upgrade program, the CMS collaboration will replace the existing forward calorimeters with 
a High Granularity Calorimeter~\cite{HGCALref}, providing a unique fine grain in view of a multi-dimensional shower reconstruction.
This is a fundamental upgrade for the whole detector given the important role of the forward calorimeter 
for physics in Phase II, it will also be crucial to complement the tracker upgrade with extended coverage in the 
forward region and a reduced material budget.

\section{The High Granularity Calorimeter}
\begin{figure}[htb!]
\centering
\includegraphics[height=3.in]{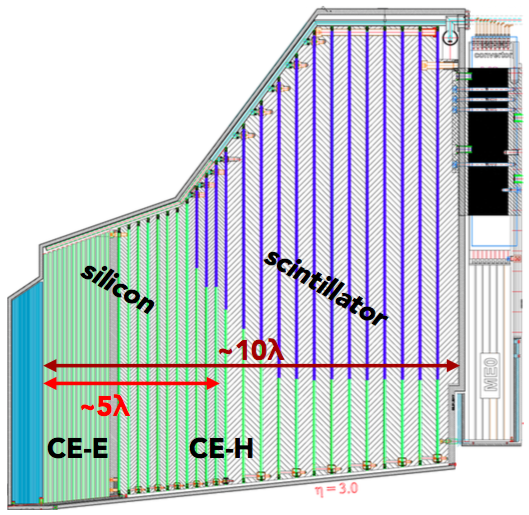}
\caption{Schematic view of the High Granularity Calorimeter design.}
\label{fig:HGCALdesign}
\end{figure}

\begin{figure}[htb!]
\centering
\includegraphics[height=2.in]{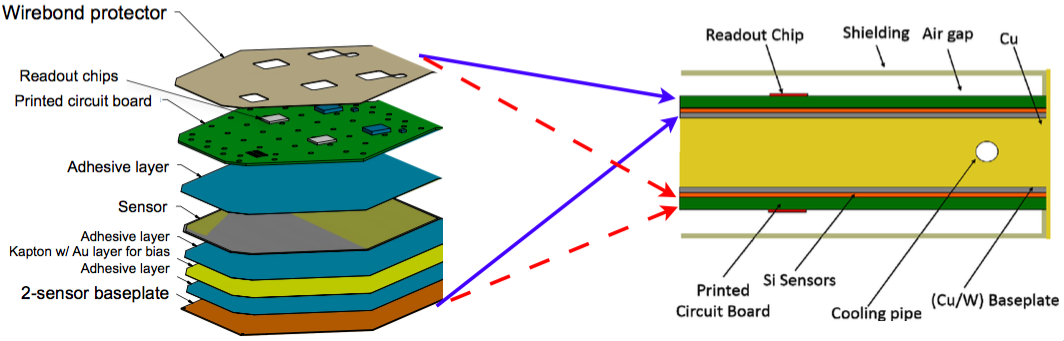}
\caption{Exploded view of an hexagon module built as a glued stack of Cu/W (25\%/75\%) baseplate, Si sensor and PCB board (on the left), 
and screwed on either side of a copper support plate (on the right). The Cu/W baseplate provides mechanical rigidity 
as well as a thermal pathway to the copper plate.
The copper plate contributes as absorber material, it provides mechanical support to the structure
and hosts the cooling pipes for removal of heat from the sensor and the readout electronics,
maintaining the detector at -30~$^{\circ}$C to limit the leakage current due to radiation.}
\label{fig:EEdesign}
\end{figure}

In the mechanical design the HGCAL consists of a sampling calorimeter with silicon and scintillators as active material, 
including both the electromagnetic (EE) and the hadronic (FH$+$BH) sections~\footnote{To reflect the design decision 
and the integration in the CMS detector nomenclature, the HGCAL is most recently referred to as CE, 
the electromagnetic section (EE) is designated CE-E, and the hadronic section (FH and BH) is CE-H.}. 
A schematic view is given in Figure~\ref{fig:HGCALdesign}.
Silicon is the main active material, it is used in the electromagnetic and innermost regions of the hadronic section, 
where the radiation is expected to be higher (up to 10$^{16}$~n/cm$^2$). It is transversely segmented into 
hexagon cells of about 1~cm$^2$ surface, for a total of over 6 million channels. 
Plastic scintillator tiles are used in the outermost regions of FH and BH.\\
In the electromagnetic part, to accommodate 28 sampling layers in about 30 cm, 
silicon sensors are mounted on either side of a copper plate, based on the stack illustrated in Figure~\ref{fig:EEdesign},
for a total of 14 copper support plates inter-spaced by lead absorbers. 
The thickness of the EE part amounts to about 25X$_{0}$ and about 1$\lambda$.
The hadronic part extends for about 1.5~m in depth and comprises 12 sampling layers in each of the FH and BH sections
with stainless steel as absorber. The thickness of the hadronic part corresponds to about 3.5$\lambda$ and 5.7$\lambda$
for the FH and BH respectively, for a total of about 9$\lambda$ for the 24 layers.

\section{Reconstruction at the HL-LHC}
The calorimeter design, with fine granularity both lateral and longitudinal, is ideally suited for particle flow reconstruction~\cite{PFrecoCMS}, 
to enhance the pattern recognition in the high congestion typical of the events at the HL-LHC.
Within the particle flow, the tracks reconstructed in the tracker are matched to 
the electromagnetic and hadronic showers individually reconstructed and identified in the calorimeter, 
where the high granularity is exploited to help in the separation of adjacent and almost overlapping particles.\\
Since the HGCAL is effectively an imaging calorimeter, a 3D imaging-clustering, 
inspired by the article in~\cite{3DclusteringGeneral}, 
is being developed to provide efficient reconstruction, resolving individual particles in the
140/200 pileup environment of the HL-LHC.
The algorithm currently proceeds in 2 steps, to first identify 2d-clusters
on each layer, based on the energy-density of the cells, and then gathering into multi-clusters 
all the 2d-clusters found aligned along the shower axis over consecutive layers.
The algorithm can be extended to more than two dimensions, to fully 
exploit the 5D potential of the calorimeter (energy, x-~y-~z-position, time).


\section{Beam tests of the EE prototypes}
Since the technical proposal (TP)~\cite{HGCALref}, submitted at the end of 2015 with a basic design of the detector,
a lot of progress has been made in the design of the mechanics and 
an extensive prototyping phase has started. 
In 2016, the first prototypes of the EE silicon hexagonal modules
were tested on the beam in campaigns organized both at FNAL and CERN, 
with the aim to give a proof of concept of the proposed design and compare the 
measured performance with a detailed simulation.

\subsection{The detector prototypes}
\begin{figure}[htb!]
\centering
\includegraphics[height=3in]{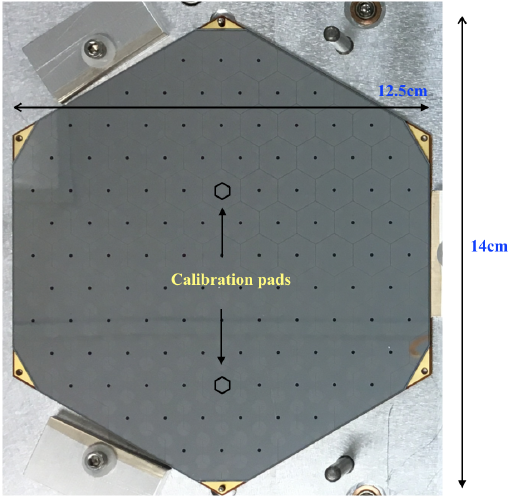}
\caption{A 6'' silicon sensor with 128 cells used in the 2016 beam tests.}
\label{fig:Hexagon}
\end{figure}

\begin{figure}[htb!]
\centering
\includegraphics[height=4.in]{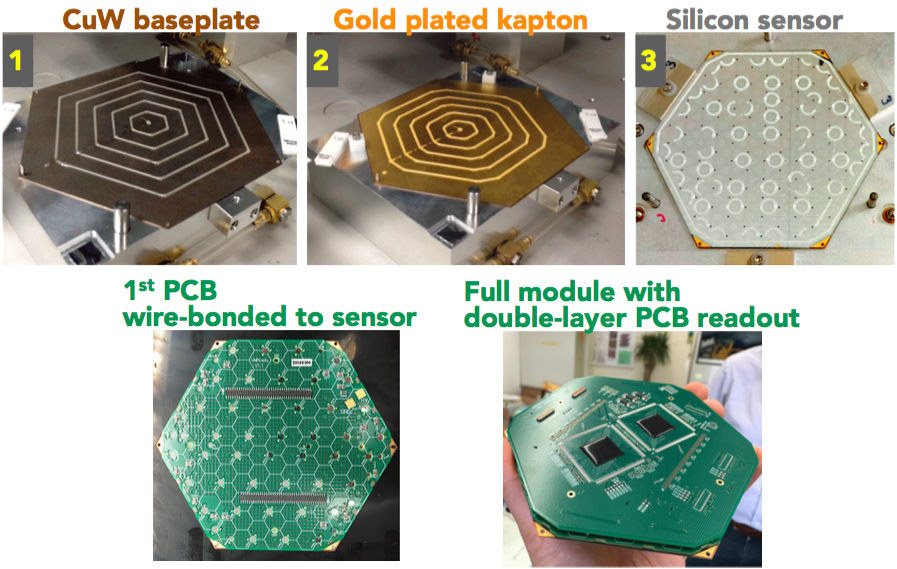}
\caption{Module assembly for the test beam prototype.}
\label{fig:moduleAssembly}
\end{figure}

\begin{figure}[htb!]
\centering
\includegraphics[height=3in]{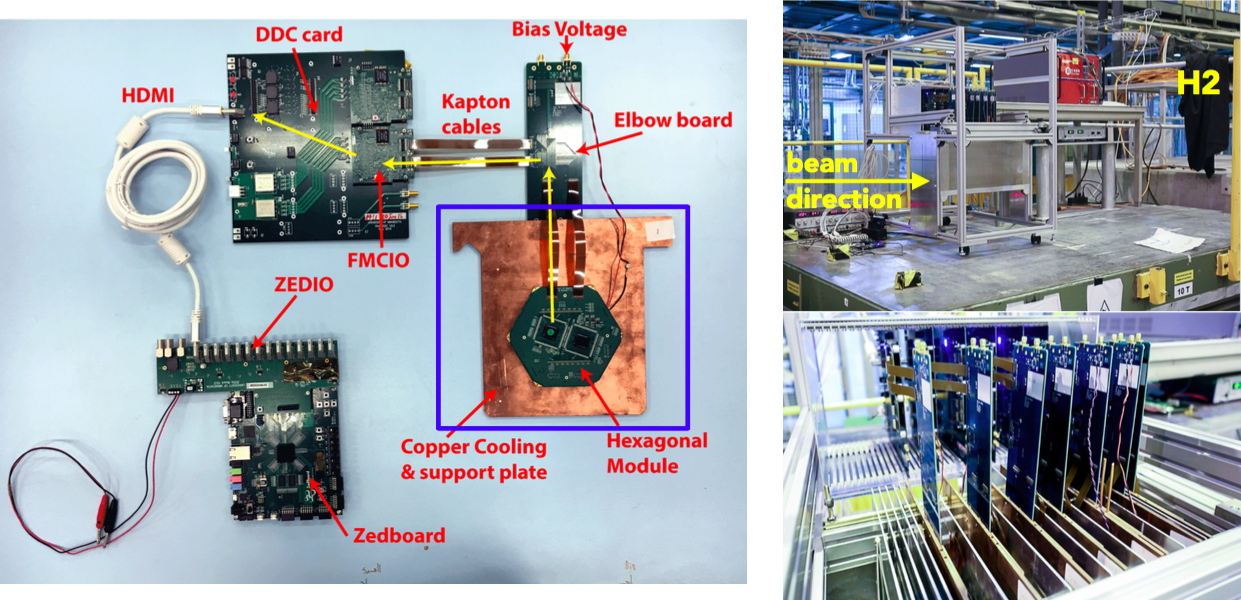}
\caption{(left) Picture of a full module assembled and mounted on a copper support plate. The full DAQ chain 
made of cables that carry data, control and low voltage along with the high voltage connections are also visible.
(right) Six copper plates with 8 silicon sensors mounted (2 copper plates are double sided) 
and fully instrumented are inserted in the hanging file system. Hanging layers of lead absorbers are also visible.}
\label{fig:hangingFile}
\end{figure}

The prototypes used for the beam tests made use of active layers and absorbers built in accordance to the TP design.
A support system consisting of an hanging file structure was used to insert active and absorber layers 
so to obtain a sampling calorimeter prototype.\\
The silicon sensors were made from 6'' wafers, of 200$\mu$m depleted region, manufactured by HPK~\footnote{Hamamatsu Photonics, Hamamatsu, Japan}, 
and cut into hexagonal shape. A picture of a silicon sensor used is visible in Figure~\ref{fig:Hexagon}.
The active modules were assembled as a glued stack of hexagonal components,  
as illustrated in Figure~\ref{fig:moduleAssembly}.
The full module was then screwed to a copper plate properly shaped for insertion in the hanging file system.
In Figure~\ref{fig:hangingFile} a picture of the fully instrumented module is shown 
together with a sampling prototype tested at CERN.\\
By exploiting the flexibility of the hanging file design, several sampling configurations were tested,
despite the limited number of available elements:
at FNAL, a 16 layer calorimeter, sampling the shower from 0.6~X$_0$ to 15.3~X$_0$, 
while at CERN two different configurations of 8 layers each, 
to measure the core (6~to~15~X$_0$) and the tail of the shower (5~to~27~X$_0$).

\subsection{Response to MIP and electron showers}
\begin{figure}[htb!]
\centering
\includegraphics[height=2.in]{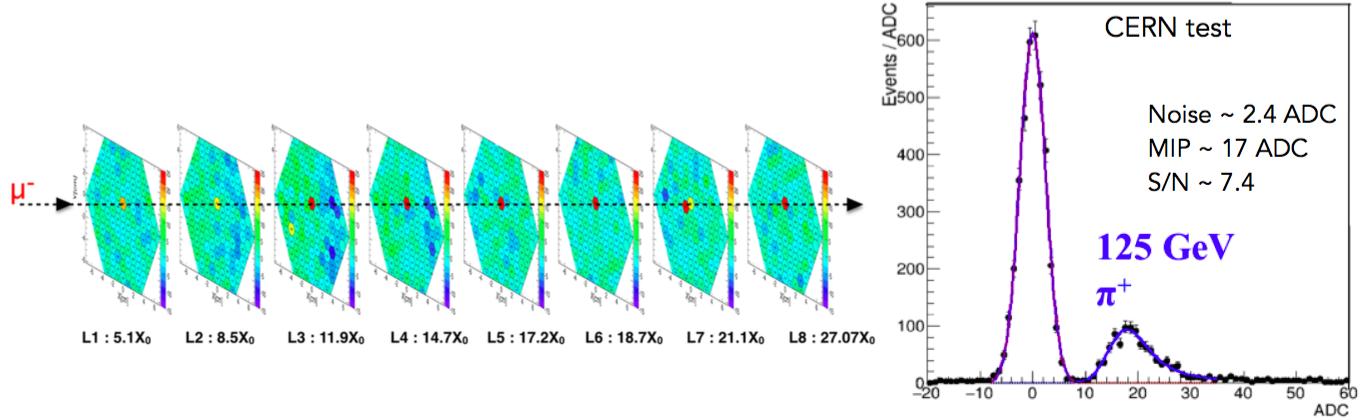}
\caption{(left) Event display of a MIP through the stack of 8 layers.
(right) ADC-count distribution for the central channel in the first layer, with the pion beam.}
\label{fig:responseToMIP}
\end{figure}

To calibrate the response between the cells, the modules were exposed to a beam
of 125~GeV pions and 120~GeV protons, at CERN and FNAL respectively.
In Figure~\ref{fig:responseToMIP} the event display of a single minimum-ionizing-particle
passing through a stack of 8 layers is shown, together with the typical spectrum 
of the energy deposition of a MIP. By comparing the most probable value obtained for the MIP energy deposition
to the noise, a S/N $\simeq$ 7 is measured which is in good agreement with the expectation.

\begin{figure}[htb!]
\centering
\includegraphics[height=2.in]{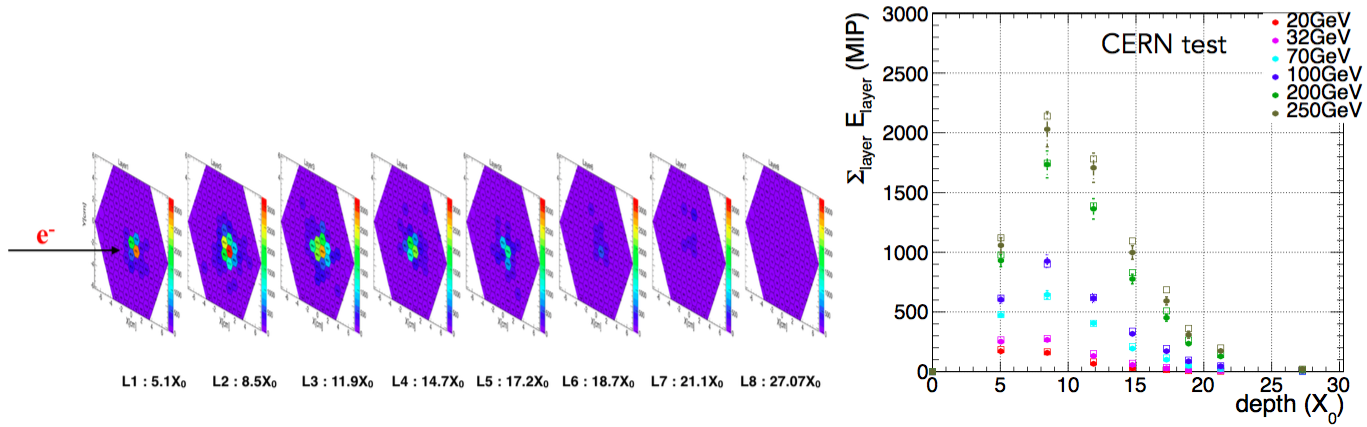}
\caption{(left) Event display of a 250~GeV electron passing through the 8 layers sampling from $\sim$5~X$_0$ to $\sim$27~X$_0$
in the CERN test. (right) Longitudinal shower profile measured in data and compared to simulation as a function of the depth in the shower.}
\label{fig:responseToElectrons}
\end{figure}

The response of the modules to electromagnetic showers was measured
by exposing the prototypes to an electron beam in the energy range [20-250]~GeV and [4-20]~GeV, 
at CERN and FNAL respectively.
In Figure~\ref{fig:responseToElectrons} the shower evolution of a 250~GeV electron 
is shown in the 8 layers setup, when sampling the shower from $\sim$5~X$_0$ to $\sim$27~X$_0$.
In the same figure the total energy reconstructed over each layer is plotted as a function of the depth in the shower, 
for all the energy points tested at CERN and it is compared to the expectation from simulation.
For the energy reconstruction, a raw calibration to account for the energy lost in the absorber material 
is applied.
As it can be noted, the shower initiated within the first 5~X$_0$ is not sampled with this specific configuration tested at CERN, 
while it was measured in the complementary configuration with 16 layers tested at FNAL.

\subsection{Results}
\begin{figure}[htb!]
\centering
\includegraphics[height=4.in]{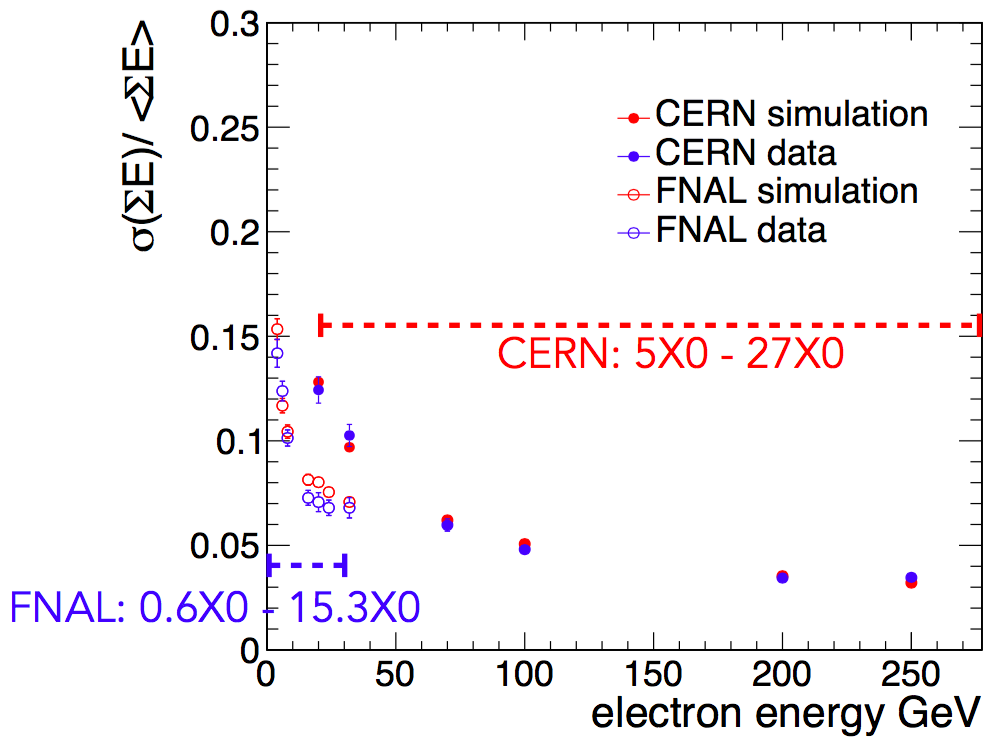}
\caption{Relative energy resolution measured as a function of the electron energy in data and simulation, 
for tests at FNAL and CERN.}
\label{fig:resolution}
\end{figure}

Data measured at the beam test were compared to a dedicated simulation, 
where the geometry of the beam line and the detector components were carefully described
and the analysis procedure and calibration were applied consistently on the two datasets.
Shower shapes variables, such as the transverse shower profiles and the fraction of energy measured in the first layer, 
were compared between data and simulation. The good agreement found is 
an indication of the reliability of the showering model used in the simulation as well as of the accurate description of the upstream material.\\
As a representative result of the analysis, in Figure ~\ref{fig:resolution}, 
the relative energy resolution measured is shown 
as a function of the beam energy for both data and simulation, 
showing a good agreement between the two.
Results from the test at FNAL and those obtained at CERN are displayed on the same canvas 
to emphasize the different sampling regimes of the two setups.
The limit in the longitudinal sampling clearly limits the possible electron energy resolution achievable, 
which is here found at the level of few percents at the highest energies, 
whereas it is expected to be close to 1~\% at 300~GeV for the final calorimeter, 
where a finer sampling ($\sim$1~X$_{0}$) will be provided.

\section{Conclusions}
The High Granularity Calorimeter for the HL-LHC is a very ambitious project, 
with an unprecedented granularity level it offers an high potential in the reconstruction to 
exploit the combined information of time, position and pulse-height to disentangle
the very complex events that we will see at the HL-LHC.
With the Technical Design Report targeted by the end of 2017, it is a critical moment for the detector design, 
with a main review taking place.
The beam test campaign of 2016, to validate the proposed design for the EE silicon modules
and provide a basic validation of the simulation 
was a fundamental step towards the consolidation of the design and
the validation of the reconstruction algorithms.
The aim of the 2017 campaign is to test an extended prototype including
electromagnetic and hadronic sampling layers, to measure the response to hadronic showers.
Tests and data analysis are ongoing at the time of writing this proceeding.

\end{document}